\documentclass[12pt]{cimtec}  %>>> use for US letter paper

\usepackage{graphicx}
\graphicspath{{Figures/}}
\DeclareGraphicsExtensions{.eps,.ps}

\usepackage{amsmath}
\usepackage{amsfonts}
\usepackage{amssymb}

\usepackage{indentfirst}

\title{\small NONLINEAR DYNAMICS OF METALLIC NANOFABRICATION}

\author{J\'er\^ome~B\"URKI
\skiplinehalf
Department of Physics, University of Arizona, 1118 E. 4th St., Tucson, AZ, USA%
\footnote{This work was supported through grant SFB-276 of the deutsche Forschungsgemeinschaft and NSF Grant No.\ DMR0312028}
}

\begin{document}

\maketitle

%%%%%%%%%%%%%%%%%%%%%%%%%%%%%%%%%%%%%%%%%%%%%%%%%%%%%%%%%%%%%
\begin{abstract}%
Using concepts from fluid dynamics, a partial differential equation for the shape evolution of a metallic nanowire is derived from a semiclassical energy functional that includes
electron-shell effects. %\ \cite{burki03}.
A rich dynamics, involving movement and interaction of kinks connecting locally stable radii, leads to the formation of a wire whose equilibrium shape is universal and consists of a
cylindrical part connected to unduloid-like leads. The universality of the equilibrium shape may provide an explanation for the formation of cylindrical nanowires observed in recent experiments. %by Kondo and Takayanagi.%\cite{kondo00}
\end{abstract}

%\keywords{Nanowires, Lifetime, Thermal fluctuations}
\renewcommand{\baselinestretch}{1.5}
\normalsize

%%%%%%%%%%%%%%%%%%%%%%%%%%%%%%%%%%%%%%%%%%%%%%%%%%%%%%%%%%%%%
\section{INTRODUCTION}
\label{sec:intro}
%%%%%%%%%%%%%%%%%%%%%%%%%%%%%%%%%%%%%%%%%%%%%%%%%%%%%%%%%%%%%

Recent transmission electron microscopy (TEM) experiments by the groups of  Takayanagi\cite{kondo97,kondo00,oshima03,oshima03a} and Ugarte\cite{rodrigues00,rodrigues02b} have found that long, almost perfectly cylindrical Au and Ag nanowires are frequently formed and have lifetimes of the order of seconds at room temperature.
In those experiments, holes are burned through a thin metallic film using high-intensity electron beams until two holes come close together, leaving a thin, initially rough nanowire in-between. After reduction of the electron irradiation to turn into imaging mode, the wires are found to evolve towards almost perfect, long cylinders. Subsequent thinning of the wire is found to occur through motion of a ``kink'' from one end of the wire to the other\cite{oshima03}.

The stability of long cylindrical wires might seem puzzling since nanowires have a large surface to volume ratio, suggesting that a wire longer than its circumference should break up under surface tension (Rayleigh instability\cite{chand81}).
This apparent paradox is resolved theoretically\cite{kassubek00,zhang02a,urban03} with the inclusion of quantum effects, which have been shown to stabilize the wires up to large temperatures for a set of ``magic'' radii. The stability arises through a
competition of shell-effects, comparable to what happens in metal clusters,
and an interplay of Rayleigh and Peierls instabilities\cite{urban03}.

A dynamical model\cite{burki03}, including surface self-diffusion of a continuous ionic background, indicates that cylindrical wires form spontaneously from random initial wires.
In this paper, the dynamical model is described and applied to the evolution of initially random nanowires, emphasizing striking similarities between simulations and available experimental results.

%%%%%%%%%%%%%%%%%%%%%%%%%%%%%%%%%%%%%%%%%%%%%%%%%%%%%%%%%%%%%
\section{MODEL}
\label{sec:model}
%%%%%%%%%%%%%%%%%%%%%%%%%%%%%%%%%%%%%%%%%%%%%%%%%%%%%%%%%%%%%

We use a model of a nanowire consisting of free electrons confined to an
axisymmetric wire by a hard-wall potential \cite{stafford97a}.  A
wire parallel to the $z$-axis is characterized by its radius $R(z,t)$ in the interval $[0,L]$,
and periodic boundary conditions \cite{com.leads} are used to extend $R(z)$.
The restriction to axisymmetric wires, though it ignores the stability of a few Jahn-Teller distorted wires at low conductance\cite{urban04}, is not expected to alter significantly the dynamics of nanowires, since the most stable nanowires have been shown to be cylinders, and axisymmetry, once present, is preserved by the diffusion equation, so that
Jahn-Teller distortions are most likely suppressed dynamically.
The free-electron model requires good screening and a spherical Fermi surface, both conditions which are met by alkali metals, such as sodium, and to some extent by noble metals.
Throughout this paper, we use parameters corresponding to Na, but results should be qualitatively unchanged for Au and Ag, which are more frequently used in imaging experiments.

In the spirit of the Born-Oppenheimer approximation, the total energy
of the nanowire is taken to be the electronic energy.  Since we are
dealing with an open system of electrons, the grand-canonical
potential is used, and can be separated into Weyl and mesoscopic
contributions \cite{stafford99,brack97}.  Dropping the
volume contribution, assumed to be constant so that the volume per atom is fixed, the energy functional is
\begin{equation}\label{eq:omega}  %%%  Eq. 1  %%%%%%%%%%%%%%%%%%%%%%%%%
%%%%%%%%%%%%%%%%%%%%%%%%%%%%%%%%%%%%%%%%%%%%%%%%%%%%%%%%%%%%%%%%%%%%%%%
\Omega\bigl[T,R(z,t)\bigr]=\sigma(T){\cal S}\bigl[R(z,t)\bigr]+\int_0^L
V\bigl(T,R(z,t)\bigr)\text{d}z,
\end{equation}  %%%%%%%%%%%%%%%%%%%%%%%%%%%%%%%%%%%%%%%%%%%%%%%%%%%%%%%
where $\sigma=\frac{\varepsilon_Fk_F^2}{80\pi}$ is the surface tension\cite{com.sigma}, $\varepsilon_F$ and $k_F$ are respectively the Fermi energy and wavevector, ${\cal S}$
is the surface area of the wire, and $V$ is a mesoscopic electron-shell potential.  Higher-order terms\cite{stafford99,brack97} proportional to the mean curvature, etc.,
are neglected.
The electron-shell potential $V$ can be expressed in terms of a Gutzwiller-type trace formula\cite{burki03}
\begin{equation}\label{eq:gutzwiller}  %%% Eq. 2  %%%%%%%%%%%%%%%%%%%%%
%%%%%%%%%%%%%%%%%%%%%%%%%%%%%%%%%%%%%%%%%%%%%%%%%%%%%%%%%%%%%%%%%%%%%%%
V(T,R) = \frac{2\varepsilon_F}{\pi}
\sum_{w=1}^{\infty}\sum_{v=2w}^{\infty} a_{vw}(T)
\frac{f_{vw}\cos\theta_{vw}}{v^2L_{vw}},
\end{equation}  %%%%%%%%%%%%%%%%%%%%%%%%%%%%%%%%%%%%%%%%%%%%%%%%%%%%%%%
where the sum includes all classical periodic orbits $(v,w)$ in a disk
billiard \cite{brack97}, characterized by their number of vertices $v$
and winding number $w$, $L_{vw}=2vR\sin(\pi w/v)$ is the length of
orbit $(v,w)$, and $\theta_{vw}=k_FL_{vw}-3v\pi/2$.  The factor
$f_{vw}=1\text{ for } v=2w$, $f_{vw}=2$ otherwise, accounts for the invariance
under time-reversal symmetry of some orbits, and $a_{vw}(T) =
\tau_{vw}/\sinh{\tau_{vw}}$ ($\tau_{vw}=\pi k_FL_{vw}T/2T_F$, $T_F$ being the Fermi temperature) is a temperature-dependent damping factor.
A semiclassical formula similar to Eq.\ (\ref{eq:gutzwiller}) was obtained for perturbations of a cylinder\cite{kassubek01}, but we point out that Eq.\ (\ref{eq:gutzwiller}) remains valid for large deformations, as long as new classes of non-planar orbits can be neglected (adiabatic approximation)\cite{burki03}.

\begin{figure}[ht]
\begin{center}\vspace*{0.7cm}
%%%%%%%%%%%%%%%%%%%%%%%%%%%%%%%%%%%%%%%%%%%%%%%%%%%%%%%%%%%
\includegraphics[width=12cm]{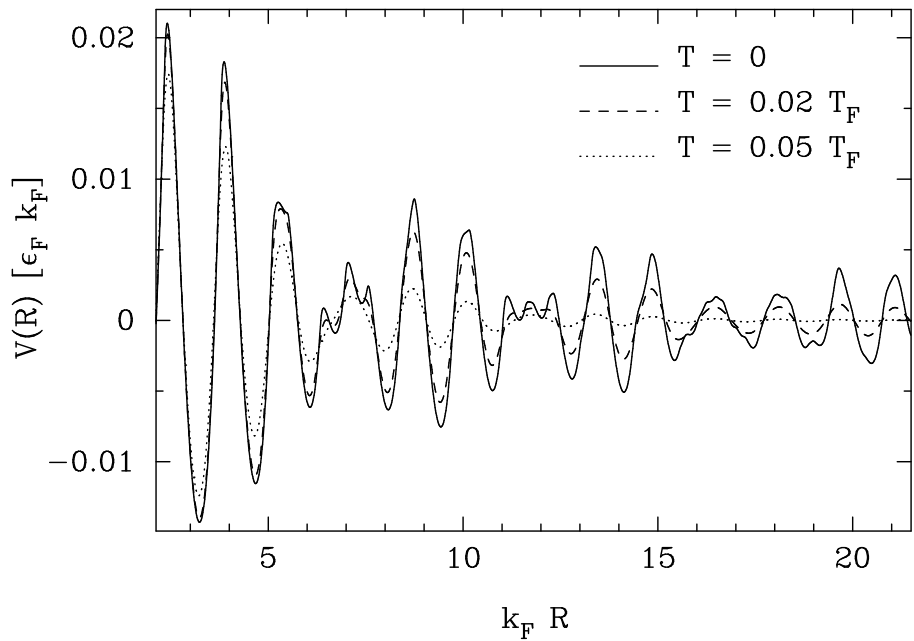}  %{Potential}
\vspace*{-0.3cm}

\caption{Mesoscopic electron-shell potential $V(R,T)$, for three different temperatures.}
\label{fig:potential}
\end{center}\end{figure}

Eqs.\ (\ref{eq:omega}) and (\ref{eq:gutzwiller}) define an energy
functional which is general and simple enough to solve for nontrivial
nanowire geometries with a wide range of radii.  The multiple deep
minima of $V(R)$, shown in Fig.\ \ref{fig:potential}, favor certain
magic radii \cite{yanson99} and suggest a multistable field theory
analogous to the sine-Gordon model.   Solutions consisting of cylindrical segments connected by kinks may therefore be anticipated.

The dynamical model is derived from two main assumptions:  First,
in the thin wires considered, a significant fraction of the atoms are at the surface
and thus surface self-diffusion is the dominant mechanism of ionic motion.
Second, under the Born-Oppenheimer approximation, the electronic
energy (\ref{eq:omega}) acts as a potential for the ions, and the
dynamics derive from ionic mass conservation:
\begin{equation}\label{eq:diffusion}  %%%  Eq. 3  %%%%%%%%%%%%%%%%%%%%%
%%%%%%%%%%%%%%%%%%%%%%%%%%%%%%%%%%%%%%%%%%%%%%%%%%%%%%%%%%%%%%%%%%%%%%%
\frac{\partial R(z,t)}{\partial t} =
-\frac{v_a}{R(z,t)}\frac{\partial}{\partial
z}\bigl[R(z,t)J_z(z,t)\bigr],
\end{equation}  %%%%%%%%%%%%%%%%%%%%%%%%%%%%%%%%%%%%%%%%%%%%%%%%%%%%%%%
where $v_a=3\pi^2/k_F^3$ is the volume of an atom, and the
$z$-component of the surface current is given by Fick's law:
\begin{equation}\label{eq:current} %%%  Eq. 4  %%%%%%%%%%%%%%%%%%%%%%%%
%%%%%%%%%%%%%%%%%%%%%%%%%%%%%%%%%%%%%%%%%%%%%%%%%%%%%%%%%%%%%%%%%%%%%%%
J_z = -\frac{\rho_SD_S}{k_B T}\frac{1}{\sqrt{1+(R')^2}}
\frac{\partial\mu}{\partial z},
\end{equation}  %%%%%%%%%%%%%%%%%%%%%%%%%%%%%%%%%%%%%%%%%%%%%%%%%%%%%%%
where $R'=\partial R/\partial z$, and $\rho_S$ and $D_S$ are the
surface density of ions and the surface self-diffusion coefficient,
respectively.  The precise value of $D_S$ for alkali metals is not
known, but it can be removed from the evolution equation by rescaling
time to the dimensionless variable $\tau=(\rho_SD_ST_F/T)t$.  The
chemical potential $\mu$,  calculated as the change in the
energy (\ref{eq:omega}) with the addition of an atom at point $z$, is
\begin{multline}\label{eq:mu}  %%%  Eq. 5  %%%%%%%%%%%%%%%%%%%%%%%%%%%%
%%%%%%%%%%%%%%%%%%%%%%%%%%%%%%%%%%%%%%%%%%%%%%%%%%%%%%%%%%%%%%%%%%%%%%%
\mu(z) =
-\frac{2\varepsilon_F}{5}+\frac{v_a\sigma}{R\sqrt{1+{R'}^2}}
\Bigl(1-\frac{RR''}{1+{R'}^2}\Bigr) %\\
-\frac{3\varepsilon_F}{k_F^2R^2}\sum_{w,v}
\frac{a_{vw}f_{vw}}{v^2}\Biggl[\sin\theta_{vw}+
b_{vw}\frac{\cos\theta_{vw}}{k_FL_{vw}}\Biggr],
\end{multline}  %%%%%%%%%%%%%%%%%%%%%%%%%%%%%%%%%%%%%%%%%%%%%%%%%%%%%%%
where $b_{vw}(T) = a_{vw}\cosh\tau_{vw}$.  Eq.\ (\ref{eq:diffusion})
is solved numerically using an implicit scheme.

This model, though neglecting the atomic structure of the wires, is expected to give reasonable insights into the dynamics and equilibrium shape of metallic nanowires: Delocalized electrons determine the cohesion of metals, so that the confinement of the electrons to the wire should affect the geometry of the wire in a significant manner.
To lowest order, atoms may be expected to arrange themselves in order to fit in the shape determined by the electrons.

\newlength\figwidth
\setlength\figwidth{12.5cm}
\begin{figure}[t]
\begin{center}
\includegraphics[width=\figwidth]{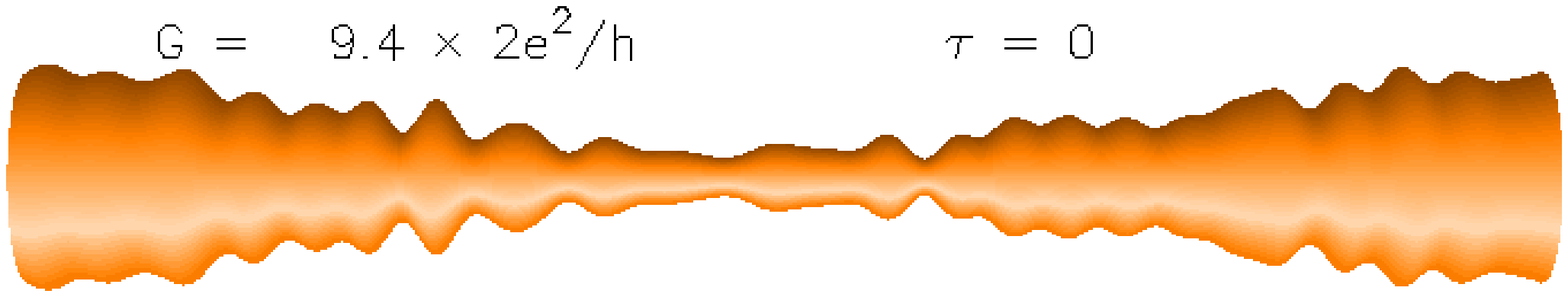} \\[0.5cm]  %{random-00} \\
\includegraphics[width=\figwidth]{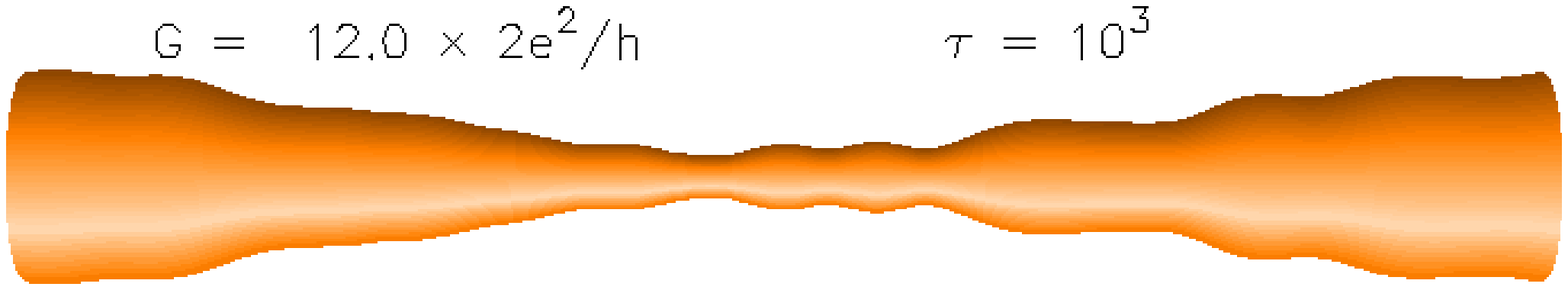} \\[0.5cm]  %{random-03} \\
\includegraphics[width=\figwidth]{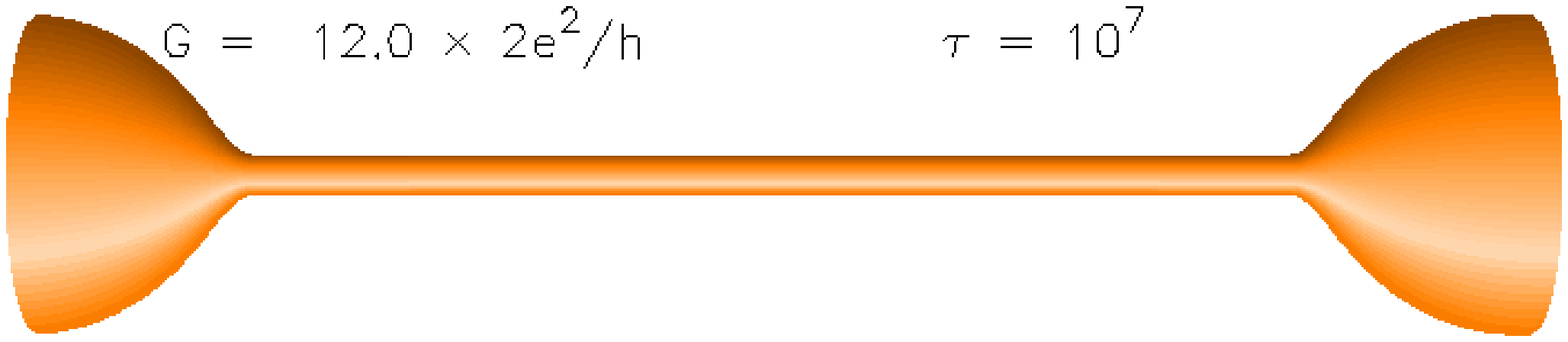}     %{random-07}
\caption{Evolution of an initially (top) random wire of length $k_FL=600$. Bottom wire shows the equilibrium shape, while the middle wire shows an intermediate stage of evolution. The conductance and dimensionless time of evolution are indicated above each wire.}
\label{fig:random}
\end{center}
\end{figure}
\section{EVOLUTION OF A RANDOM WIRE}
\label{sec:evol}
%%%%%%%%%%%%%%%%%%%%%%%%%%%%%%%%%%%%%%%%%%%%%%%%%%%%%%%%%%%%%

%
\begin{figure}[b]
\begin{center}
\setlength{\unitlength}{1mm}
\begin{picture}(140,87)(0,0)
\put(0,0){\includegraphics[width=12cm]{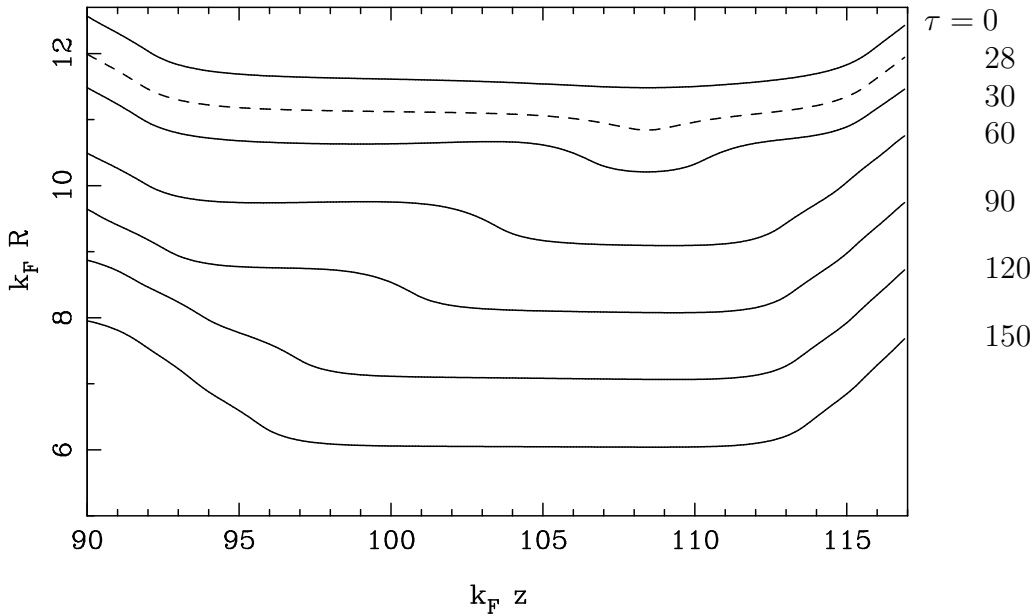}}   %{Kink2}}
\put(122,78){$\tau = 0$}
\put(130,73){$28$}
\put(130,68){$30$}
\put(130,63){$60$}
\put(130,54){$90$}
\put(130,45){$120$}
\put(130,36){$150$}
\end{picture}
\caption{Thinning of a wire through nucleation and subsequent motion of a kink.
From top to bottom, the solid lines show $R(z,1960+\tau)$ for part of the wire at times indicated on the right of each line (curves for $\tau<150$ shifted vertically for clarity), while the dashed line shows an intermediate step of evolution during the nucleation of the kink/antikink pair.}
\label{fig:kink}
\end{center}
\end{figure}
In order to mimic TEM experiments\cite{kondo97,kondo00,oshima03,oshima03a,rodrigues00,rodrigues02b}, we perform simulations starting from an initially random wire given by
\begin{equation}
R(z,t=0)=\sum_{|q|<q_{max}} R_q e^{\text{i} q z},
\label{eq:random}
\end{equation}
where $q=2\pi n/L$ with $n$ an integer, $q_{max}$ is of the order of the Fermi wavevector $k_F$, and $R_q^*=R_{-q}$ to ensure the reality of $R(z)$. The coefficients $R_0$ and $R_{\pm 1}$ are chosen so as to set the initial average radius and curvature of the wire, while other coefficients are random.
We performed 14 simulations, starting with different random wires of various radii and lengths.
Figure\ \ref{fig:random} shows such a wire, whose evolution is typical, at three stages of evolution: {\sl (i)} Initial random shape; {\sl (ii)} After a relatively short time, the short
wavelength roughness has been smoothed out, leaving a few cylindrical segments, connected by kinks;
{\sl (iii)} Eventually, all kinks move towards the ``leads'', yielding an equilibrium shape consisting of a cylindrical part, whose radius and length depend on the initial conditions, and thicker contacts, whose shape will be discussed in the next section.
This suggests that the natural evolution of a nanowire, at a temperature sufficient for atoms to diffuse, is to form a cylinder, providing an explanation of the TEM observation. %Takayanagi's group\cite{kondo97,kondo00}.

Figure\ \ref{fig:kink} shows details of the thinning process:
 The top solid line shows the central part of a wire, with a conductance $G=8\,G_0$, right before the thinning starts with the nucleation of a kink/antikink pair close to one end of the wire (dashed line). One of the kinks is quickly absorbed into the closest lead, while the other moves at constant speed along the wire towards the other lead, where it is also absorbed. The final wire (bottom curve) has a conductance of $G=6\,G_0$.
Similar thinning by motion of a kink has been reported in TEM experiments\cite{rodrigues02b,oshima03}.

\setlength\figwidth{13.4cm}
\newlength\figwidthp
\setlength\figwidthp\figwidth
\addtolength\figwidthp{0.2cm}
\begin{figure}[t]
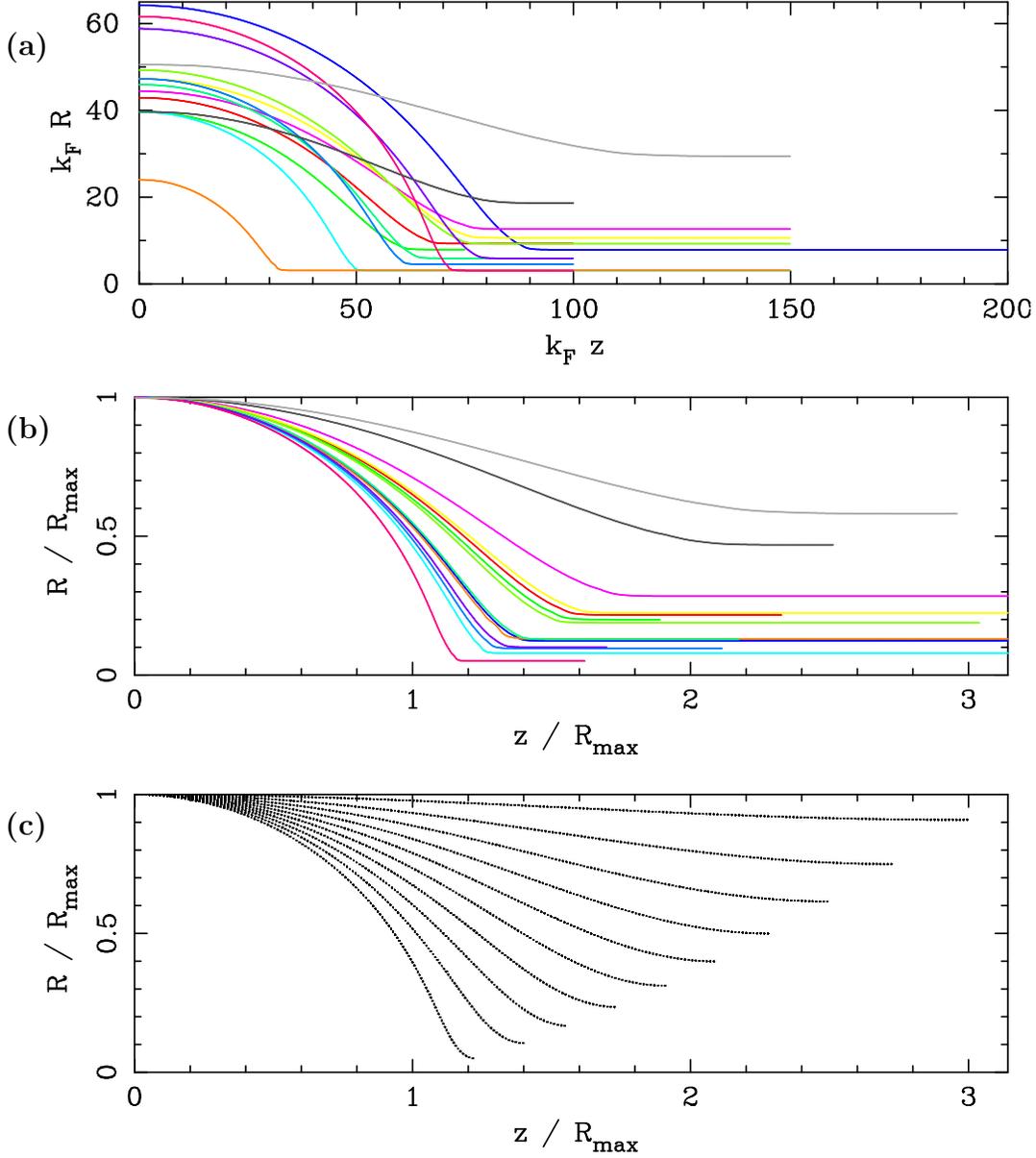

\begin{center}
\setlength{\unitlength}{1mm}
\begin{picture}(142,163)(0,0)
\put(1,109){\includegraphics[width=\figwidthp]{Figure4a}}  %{Wires}}
\put(-5,152){\text{\bf(a)}}
\put(0,55){\includegraphics[width=\figwidth]{Figure4b}}    %{Wires-scaled}}
\put(-5,99){\text{\bf(b)}}
\put(0,0){\includegraphics[width=\figwidth]{Figure4c}}     %{Unduloids}}
\put(-5,44){\text{\bf(c)}}
\end{picture}
%\hspace*{0.2cm}\includegraphics[width=\figwidthp]{Wires} \\
%\includegraphics[width=\figwidth]{Wires-scaled}
\caption{(a) Equilibrium shape for 14 different initial random wires, with conductances ranging from 1 to 200 $G_0$, and length $200 \leqslant k_FL \leqslant 600$; (b) Same wires rescaled by their maximum radius $R_{max}$; (c) Delaunay unduloids of various curvature.}
\label{fig:universal}
\end{center}
\end{figure}
\section{UNIVERSALITY OF THE EQUILIBRIUM SHAPE}
\label{sec:universal}
%%%%%%%%%%%%%%%%%%%%%%%%%%%%%%%%%%%%%%%%%%%%%%%%%%%%%%%%%%%%%

Radius functions $R(z,t\rightarrow\infty)$ in the interval $[0,L/2]$ for equilibrium shapes obtained from 14 simulations of random wires are shown in Fig.\ \ref{fig:universal}(a). The conductance of those wires ranges from $1$ to $200\,G_0$ with lengths $200 \leqslant k_FL \leqslant 600$. They all have in common a cylindrical part with a spherical-like lead. Fig.\ \ref{fig:universal}(b) shows that rescaling all lengths by the maximum radius $R_{max}$ of each wire reveals a regularity of the shape of the lead: It is actually a close approximation of a Delaunay unduloid\cite{bernoff98}, whose curvature is determined solely by the ratio of the radius of the cylindrical part and $R_{max}$. Figure\ \ref{fig:universal}(c) shows several unduloids of various curvature values, to be compared with the rescaled shapes of Fig.\ \ref{fig:universal}(b). The Delaunay unduloid of revolution is a surface of constant mean curvature and is an unstable steady state of diffusion equation (\ref{eq:diffusion}) without the shell-effect term. In our case, the deep minima of the electron-shell potential, Fig.\ \ref{fig:potential}, pin the unduloid at its connection with the cylindrical part, thus stabilizing it. In fact, if one switches off the electron-shell potential in the simulations, the equilibrated wires break apart, as expected from the Rayleigh instability.

\section{CONCLUSIONS}
\label{sec:conclusion}
%%%%%%%%%%%%%%%%%%%%%%%%%%%%%%%%%%%%%%%%%%%%%%%%%%%%%%%%%%%%%%%%%%%%%%%%

We have shown that a dynamical model based on the nanoscale free-electron model\cite{stafford97a} and including surface self-diffusion of a continuous ionic background accounts for the observed formation of nearly perfect cylindrical wires in TEM experiments. The thinning of the wire appears to occur in very similar ways in both experiments and  simulations: The wire necks down near one lead, leaving a kink, or step, that moves towards the other lead at more or less constant speed.
We show that the shape of the equilibrium wire is universal, and consists of a cylinder connected to unduloid-like leads.

\acknowledgments
I would like to thank C.~A.~Stafford and R.~E.~Goldstein for their contribution to this work.

%%%%%%%%%%%%%%%%%%%%%%%%%%%%%%%%%%%%%%%%%%%%%%%%%%%%%%%%%%%%%
%%%%% References %%%%%

\bibliography{refs}

\begin{thebibliography}{10}

\bibitem{kondo97}
\textsc{Y.~Kondo and K.~Takayanagi}, Phys. Rev. Lett. {\bf 79} (1997) 3455.

\bibitem{kondo00}
\textsc{Y.~Kondo and K.~Takayanagi}, Science {\bf 289} (2000) 606.

\bibitem{oshima03}
\textsc{Y.~Oshima, Y.~Kondo, and K.~Takayanagi}, J. Electron Microsc. {\bf 52}
  (2003) 49.

\bibitem{oshima03a}
\textsc{Y.~Oshima, A.~Onga, and K.~Takayanagi}, Phys. Rev. Lett. {\bf 91}
  (2003) 205503.

\bibitem{rodrigues00}
\textsc{V.~Rodrigues, T.~Fuhrer, and D.~Ugarte}, Phys. Rev. Lett. {\bf 85}
  (2000) 4124.

\bibitem{rodrigues02b}
\textsc{V.~Rodrigues, J.~Bettini, A.~R. Rocha, L.~G.~C. Rega, and D.~Ugarte},
  Phys. Rev. B {\bf 65}(15) (2002) 153402.

\bibitem{chand81}
\textsc{S.~Chandrasekhar}, Hydrodynamic and Hydromagnetic Stability (Dover, New
  York ).

\bibitem{kassubek00}
\textsc{F.~Kassubek, C.~A. Stafford, H.~Grabert, and R.~E. Goldstein},
  Nonlinearity {\bf 14} (2001) 167.

\bibitem{zhang02a}
\textsc{C.-H. Zhang, F.~Kassubek, and C.~A. Stafford}, Phys. Rev. B {\bf 68}
  (2003) 165414.

\bibitem{urban03}
\textsc{D.~F. Urban and H.~Grabert}, Phys. Rev. Lett. {\bf 91} (2003) 256803.

\bibitem{burki03}
\textsc{J.~B{\"u}rki, R.~E. Goldstein, and C.~A. Stafford}, Phys. Rev. Lett.
  {\bf 91} (2003) 254501.

\bibitem{stafford97a}
\textsc{C.~A. Stafford, D.~Baeriswyl, and J.~B{\"u}rki}, Phys. Rev. Lett. {\bf
  79} (1997) 2863.

\bibitem{com.leads}
The locality of the energy functional\ (\ref{eq:omega}) suppresses the
  dependence on boundary conditions.

\bibitem{urban04}
\textsc{D.~Urban, J.~B\"urki, C.-H. Zhang, C.~A. Stafford, and H.~Grabert},
  cond-mat/0312517.

\bibitem{stafford99}
\textsc{C.~A. Stafford, F.~Kassubek, J.~B\"urki, and H.~Grabert}, Phys. Rev.
  Lett. {\bf 83}(23) (1999) 4836.

\bibitem{brack97}
\textsc{M.~Brack and R.~K. Bhaduri}, Semiclassical physics (Addison-Wesley ).

\bibitem{com.sigma}
For noble metals, $\sigma$ is roughly twice as large.

\bibitem{kassubek01}
\textsc{F.~Kassubek, C.~A. Stafford, H.~Grabert, and R.~E. Goldstein},
  Nonlinearity {\bf 14} (2001) 167.

\bibitem{yanson99}
\textsc{A.~I. Yanson, I.~K. Yanson, and J.~M. van Ruitenbeek}, Nature {\bf 400}
  (1999) 144.

\bibitem{bernoff98}
\textsc{A.~J. Bernoff, A.~L. Bertozzi, and T.~P. Witelski}, J. Stat. Phys. {\bf
  93} (1998) 725.

\end{thebibliography}
\bibliographystyle{cimtecbib}

\end{document}